\documentclass[aps,prl,reprint,groupedaddress]{revtex4-2}

\usepackage{amsmath}
\usepackage{graphics}
\usepackage{tikz}
\usepackage[normalem]{ulem}
\usepackage{hyperref}

\begin{document}



\title{Nonclassical Nucleation Pathways in Liquid Condensation \\Revealed by Simulation and Theory}

\author{Yijian WU}
\affiliation{Laboratoire de Physique de la Matière Condensée, CNRS, Ecole polytechnique, Institut Polytechnique de Paris, 91120 Palaiseau, France}
\author{Thomas PHILIPPE}
\email[]{thomas.philippe@polytechnique.edu}
\affiliation{Laboratoire de Physique de la Matière Condensée, CNRS, Ecole polytechnique, Institut Polytechnique de Paris, 91120 Palaiseau, France}
\author{Aymane GRAINI}
\affiliation{Laboratoire de Physique de la Matière Condensée, CNRS, Ecole polytechnique, Institut Polytechnique de Paris, 91120 Palaiseau, France}
\author{Julien LAM}
\affiliation{Université Lille, Centre National de la Recherche Scientifique, INRA, ENSCL, UMR 8207, UMET, Unité Matériaux et Transformations, 59000 Lille, France}


\begin{abstract}
Using state-of-the-art rare-event sampling simulations, we precisely characterize the nucleation of liquid droplets from a supersaturated Lennard-Jones gas and uncover a key physical feature: critical clusters nucleate with a density that differs substantially from that of the macroscopic equilibrium liquid. Our atomistic simulations also reveal a nonclassical nucleation pathway showing simultaneous growth and densification in liquid condensation. We then exploit these insights to develop a two-variable nucleation theory, in which the cluster density is allowed to vary. Our accessible model based on the capillary approximation is able to quantitatively retrieve the numerical results in nucleation rate and critical cluster properties over a large range of supersaturation. Remarkably, the two-variable model successfully captures the observed nucleation pathway. The effectiveness of this integrated numerical and theoretical framework demonstrates that the cluster density is a decisive variable in nucleation, highlighting the limitations of the single-variable description while offering a robust foundation for its refinement.
\end{abstract}


\maketitle


\noindent\textit{Introduction}\,---\,Nucleation is the triggering process of every first order transition. As a consequence, any progress in understanding its complex mechanisms can potentially impact many research fields including material engineering~\cite{Wu2022Jan,Schick2017Oct}, pharmaceuticals~\cite{AbuBakar2009Mar,Lu2010Mar,Thakore2020Feb}, and climate control~\cite{Murray2012,Zhang2012Mar,Jun2016Sep}. So far, nucleation has mostly been described through classical nucleation theory (CNT)~\cite{volmer_kinetik_1939,becker_kinetische_1935,zeldovich_theory_1942,frenkel_kinetic_1946,Turnbull_1949,feder_homogeneous_1966}. The popularity of CNT originates from the simplicity of its ingredients namely the nucleating cluster size as a unique order parameter and the capillary model~\cite{Reguera_2003,Reguera_2003_revisited_problem}. Although highly efficient for reaching a general understanding of nucleation, CNT usually fails dramatically at providing quantitative agreement in both experiments~\cite{iland_argon_2007,sinha_argon_2010} and numerical simulations~\cite{Diemand_JCP2013,yoo_monte_2001}.  

So far, most efforts dedicated to improving CNT consisted of revisiting the capillary model, including, for instance, a size dependence for the surface tension~\cite{Prestipino}. Alternatively, assuming that the whole nucleation process is driven only by the nucleating cluster size has also been highly debated since the emergence of diffuse-interface models~\cite{Oxtoby-Evans-1988,Cahn-Hilliard-1959,Granasy_1993,Ghosh-2011}. Their predictions are in fairly good agreement with microscopic simulations~\cite{Lutsko_JCP2008, Lutsko-JCP2011,Reguera_2003}, but their highly multidimensional free energy surface complicates the determination of critical cluster properties and makes the calculation of the nucleation rate very difficult. The latter has been recently attempted by Lutsko \textit{et al.} for a colloidal solution~\cite{Lutsko-JCP2024} and by Simeone \textit{et al.} within the square-gradient approximation~\cite{Simeone_PRL2023}. Such fundamental frameworks are certainly very promising but their application to practical problems remains particularly challenging.

Meanwhile, any sort of improvement of CNT must also be confirmed via concrete observations. Experimentally, a key issue remains the presence of impurities triggering heterogeneous nucleation and the lack of control over the precise location of the nucleation events. In numerical simulations, homogeneous nucleation is considered a rare event since one has to wait for thermal fluctuations to drive the spontaneous formation of the critical nucleus~\cite{Sosso2016Jun,Brukhno2008Nov,Finney2024Jan}. The so-called incubation time depends on the supersaturation level and the associated free energy barrier. While brute-force simulations~\cite{Yasuoka_Matsumoto_JCP98} have been used to measure nucleation rates at high supersaturation, isolating the critical clusters in a large scale simulation remains very challenging especially at low supersaturation. For that purpose, rare-event sampling techniques~\cite{Hartmann2013Dec,Raucci2022Feb,Bernardi2015May}, which were developed notably for characterizing chemical reactions and protein folding have been used for studying nucleation. In this context, two approaches are usually considered: (1) biasing on the energy that allows for free energy barrier crossing yet in potentially unrealistic fashion~\cite{Bussi2020Apr,Kastner2011Nov,Comer2015Jan} and (2) biasing on the trajectory space, which is realistic but requires a huge computational effort that can hardly be achieved in low supersaturation regimes~\cite{Bolhuis2002Oct,Dellago2002Jul,Allen2009Oct}.

With two novel ideas, our work revisits the well-studied problem of liquid condensation from vapor. On the one hand, we use an innovative numerical strategy based on rare-event sampling techniques to precisely measure structural features of critical clusters in atomistic simulations over a large range of supersaturation. With unprecedented detail, our simulation results support the nonclassical nucleation nature of liquid condensation process. On the other hand, leveraging the insights from our simulation results, we construct an accessible multidimensional theoretical framework that accurately predicts the critical cluster's structural features, its associated free energy barrier, and the nucleation rate. Altogether, these advances demonstrate that a minimal multidimensional description of nucleation is necessary to overcome the limitations of CNT and to reconcile theory with simulations, paving the way for a better understanding of nucleation.

\vspace{0.25cm}
\noindent\textit{Rare event sampling}\,---\,In this first part, we will describe how different molecular dynamics (MD) simulation approaches were carried out to observe the liquid condensation from the vapor phase. In particular, we focus on the Lennard-Jones system and use the corresponding units, namely energy parameter $\epsilon$, length parameter $\sigma$, Boltzmann constant $k_\mathrm{B}$, and monomer mass $m$. Herein, our objective is not to simulate the liquid condensation which can be done with brute-force simulations using large scale simulations at moderate supersaturation regimes~\cite{Diemand_JCP2013}. Instead, we wish to isolate critical nucleation clusters and generate a representative ensemble of structures located near the saddle point of the free energy barrier, enabling a precise characterization of their properties. Depending on the supersaturation regime, two different approaches are carried out. 

In the lowest supersaturation regimes (the vapor density $\rho_0<0.015\sigma^{-3}$), the cluster size is expected to be large and we therefore are able to use the seeding technique~\cite{Vega_Seeding_JCP2016} with a particular implementation that allows us to stabilize critical clusters. For the initialization, a spherical liquid droplet is artificially inserted into a gas phase. When the density and the box size are correctly chosen, the droplet spontaneously stabilizes at a particular radius and density, which are most probably different from the initial conditions. Meanwhile, the external gas density is then automatically adjusted through finite-size effects and mass conservation at a new equilibrium value $\rho_0$, for which the obtained liquid droplet is a critical nucleation cluster~\cite{Vega-Seeding-NVT}. This behavior is known as the "superstabilization" effect in confined systems~\cite{Philippe_PRE_2017,Wilhelmsen_2015,Schmelzer_2007,Schmelzer_2011,Reguera_2003,Wilhelmsen_2014_communication}. 
Ultimately, by testing different box sizes and average densities, one can stabilize liquid clusters at different supersaturation regimes. Then, an ensemble of critical clusters can be obtained upon changing the initial velocity and atomic positions. 

In the intermediate supersaturation regimes ($\rho_0 \in [0.02\sigma^{-3}, 0.04\sigma^{-3}]$), the critical cluster is too small for the seeding approach, we therefore use a set of enhanced sampling techniques combining both energy and trajectory biasing approaches. In particular, to push gas into forming a liquid droplet despite the free energy barrier, we use steering molecular dynamics~\cite{Izrailev1999,Patel2014Feb} with the coordination number as a collective variable. Next, we extract structures from the obtained energy-biased trajectories and measure their commitment probability toward both the liquid and vapor phases by running brute-force simulations with different initial velocity conditions. Finally, we initialize aimless shooting~\cite{Mullen2015Jun,Burgin2023Jan,Peters2006Aug} simulations using one of the obtained brute-force trajectories connecting the vapor and the liquid phases. This enables us to generate many connecting trajectories and extract the associated critical nuclei. 

By coupling these two approaches, we manage to obtain approximately 20 well-defined critical clusters per studied gas density that allow us to precisely characterize the nucleation event in terms of density and size. More details on the different simulation and analysis protocols can be found in Appendix.

\vspace{0.25cm}
\noindent\textit{Theory}\,---\,Our theoretical framework builds on the multivariable nucleation theory~\cite{langer_statistical_1969,alekseechkin_multivariable_2006}. To simplify its application, we extend the capillary approximation in CNT by explicitly incorporating both the size $R$ and the density 
$\rho$ of the nucleating cluster. Consequently, the work of formation is given by $\Delta\varOmega(R,\rho)=-4\mathrm{\pi}{R}^3g_n/3+4\mathrm{\pi}{R}^2\gamma$. Here, $g_n$ is the driving force for nucleation, defined as $g_n = \mathcal{F}(\rho_0)+\frac{\partial \mathcal{F}}{\partial \rho}\Big|_{\rho=\rho_0}(\rho-\rho_0)-\mathcal{F}(\rho)$, with $\mathcal{F}$ the Helmholtz free energy per unit volume~\cite{Cahn-Hilliard-1959}. The surface tension of vapor-liquid interface, $\gamma$, is inspired by the square gradient approximation (SGA), $\gamma = \kappa(\rho-\rho_0)^2$, where $\kappa$ is assumed density-independent and can be computed from the surface tension at equilibrium~\cite{Schmelzer_2007,Philippe_2011_Phil_mag,Philippe_2024,Ghosh-2011,lutsko_two-parameter_2015}. At first we ignore the dependence of surface tension with size~\cite{Prestipino}. We note that unlike in CNT, both $g_n$, and $\gamma$ are now treated as functions of the cluster density, $\rho$.

From this expression of the work of formation, we can deduce the equilibrium density function for the nuclei: $f_\mathrm{eq} = f_0\,\exp{\big(-\beta\Delta\varOmega(R,\rho)\big)}$, where $f_0$ is the distribution parameter and $\beta=1/(k_\mathrm{B}T)$ is the inverse temperature. Meanwhile, nucleation kinetics can be probed using the Fokker-Planck equation for the time-dependent cluster density function $f(R,\rho,t)$: $\frac{\partial f}{\partial t} = -\big(\frac{\partial J_R}{\partial R} + \frac{\partial J_\rho}{\partial \rho}\big)$, where the nucleation flux vector in phase space is $\boldsymbol{J} = \big(J_R, J_\rho\big)^\mathrm{T} = -\mathbf{D} \big(\frac{\partial f}{\partial R}, \frac{\partial f}{\partial \rho}\big)^\mathrm{T} + \big(\frac{\partial R}{\partial t}, \frac{\partial \rho}{\partial t}\big)^\mathrm{T}f$. Here, $\mathbf{D}$ is the diffusivity matrix in phase space. At equilibrium, the nucleation flux vanishes, leading to $\big(\frac{\partial R}{\partial t}, \frac{\partial \rho}{\partial t}\big)^\mathrm{T} = -\beta\mathbf{D} \big(\frac{\partial \Delta\varOmega}{\partial R}, \frac{\partial \Delta\varOmega}{\partial \rho}\big)^\mathrm{T}$. 

The nucleation trajectory passes through the saddle point of the work of formation, $(R_\mathrm{c},\rho_\mathrm{c})$, defining the critical cluster size and density. The corresponding work of formation is $\Delta\varOmega_\mathrm{c}$, and the Hessian matrix at the saddle point is $\mathbf{H}$. Near this saddle point, 
\begin{equation}
    \begin{pmatrix}
        \frac{\partial R}{\partial t}  \\[3pt]
        \frac{\partial \rho}{\partial t}
    \end{pmatrix}
    = -\beta\mathbf{DH}
    \begin{pmatrix}
        R-R_\mathrm{c}  \\
        \rho-\rho_\mathrm{c}
    \end{pmatrix}
    \equiv 
    -\beta\mathbf{Z}
    \begin{pmatrix}
        R-R_\mathrm{c}  \\
        \rho-\rho_\mathrm{c}
    \end{pmatrix},
\label{eq:Z}
\end{equation}
where the matrix $\mathbf{Z} \equiv \mathbf{DH}$ is introduced in the same way as Alekseechkin~\cite{alekseechkin_multivariable_2006}.  

In closing, we can retrieve the steady-state nucleation rate in the two-variable model using~\cite{alekseechkin_multivariable_2006,langer_statistical_1969}
\begin{equation}
    I=f_0\, \exp{(-\beta\Delta\varOmega_\mathrm{c})} \, \frac{|{\lambda_\mathbf{Z}}_1|}{\sqrt{|\det \mathbf{H}|}}.
\label{eq:I}
\end{equation}
Here, ${\lambda_\mathbf{Z}}_1$ is the only negative eigenvalue of $\mathbf{Z}$, and its corresponding eigenvector ${\boldsymbol{v}_\mathbf{Z}}_1$ points along the unstable direction of nucleation, which is indeed the nucleation flux direction. Unlike in CNT and previous theoretical models~\cite{Philippe_2011_Phil_mag}, we predict that the nucleation current results from a complex interplay between thermodynamics ($\mathbf{H}$) and kinetics ($\mathbf{D}$) that is encoded in $\mathbf{Z}$. At this stage, several parameters of the model remain to be obtained in practice. 

Firstly, for the distribution parameter $f_0$, neither Alekseechkin's nor Langer's analytical expression is applicable to our studies: Alekseechkin's assumption regarding the diagonal elements of $\mathbf{H}$~\cite{alekseechkin_multivariable_2006} is not universally valid, and Langer's assumption that the metastable minimum is sharp and well-isolated~\cite{langer_statistical_1969} does not hold at all in our two-variable framework. We propose to employ mass conservation in the metastable region (MS) to calculate $f_0$, 
\begin{equation}
    \int_\mathrm{MS} f_\mathrm{eq}\, q(R,\rho)\, \mathrm{d}R \,\mathrm{d}\rho = \rho_0,
\label{eq:f0_MassConservation}
\end{equation}
where $q(R,\rho)=4\mathrm{\pi}\rho R^3/3$ is the number of monomers in a cluster. We emphasize here that the determination of $f_0$ is extremely difficult in high dimensional phase space, which makes the calculation of nucleation rate hardly achieved, while our two-dimensional framework and calculating method of $f_0$ provide a simple way to obtain nucleation rate. Unlike in Ref. \cite{lutsko_two-parameter_2015}, our calculating method of $f_0$ is normalizable.

Secondly, the diffusivity matrix $\mathbf{D}$ is derived from the formulation of dynamical density functional theory (DDFT)~\cite{Marconi_2000,Archer-Evans-DDFT-2004,lutsko_communication_2011} near the saddle point. DDFT is based on the continuity equation with the assumption that the driving force for particle diffusion is proportional to the real-space diffusion constant $D_\mathrm{diff}$ and the gradient of chemical potential~\cite{Evans01041979}. Thus, mass conservation is locally considered for the growth of clusters near the saddle point. A detailed explanation for the calculation of $\mathbf{D}$ can be found in Ref.~\cite{lutsko_dynamical_2012} and Appendix. In practice, the diffusion constant is estimated from our seeding MD simulations of the early growth stages of postcritical clusters~\cite{auerPredictionAbsoluteCrystalnucleation2001}. At $T=0.8\epsilon/k_\mathrm{B}$, we find $D_\mathrm{diff} = 0.03\sigma\sqrt{\epsilon/m}$.

Thirdly, the Helmholtz free energy of homogeneous phase of Lennard-Jones system is calculated using perturbation theory~\cite{chandler_equilibrium_1970,weeks_role_1971,andersen_relationship_1971,ree_equilibrium_1976,kang_perturbation_1985} within density functional theory, following Ref.~\cite{lutsko_effect_2005}. We do not adopt the equations of state~\cite{johnson_lennard-jones_1993,kolafa_lennard-jones_1994} since the perturbation theory is a more physically derived model and its predicted equilibrium densities ($\rho^\mathrm{eq}_1 = 0.0063\sigma^{-3}$, $\rho^\mathrm{eq}_2 = 0.7847\sigma^{-3}$) are in very good agreement with microscopic simulations at $T=0.8\epsilon/k_\mathrm{B}$. 

Finally, the surface energy factor at this temperature is determined from the surface tension at equilibrium ($\gamma^\mathrm{eq} = 0.8633\epsilon/\sigma^{2}$)~\cite{baidakov_metastable_2007} using $\kappa = \gamma^\mathrm{eq}/(\rho^\mathrm{eq}_1-\rho^\mathrm{eq}_2)^2$.

\vspace{0.1cm}
For comparison purposes, the results of our model will be compared with both CNT and the SGA diffuse-interface model, where the work of formation is a functional of the spherical density profile $\widetilde{\rho} (r)$~\cite{Cahn-Hilliard-1959,Oxtoby-Evans-1988,Granasy_1993,Ghosh-2011,Philippe_JCP_2011,Lutsko-JCP2011}, 
\begin{equation}
\label{SGA}
    \Delta \varOmega^\mathrm{SGA} = 4\mathrm{\pi} \int_0^\infty \Big(-g_n \big(\widetilde{\rho} (r)\big) + \frac{K}{2} \big( \frac{\partial \widetilde{\rho} (r)}{\partial r}\big)^2 \Big) r^2 \mathrm{d}r,
\end{equation}
with $K$ determined from the surface energy of the planar interface at equilibrium~\cite{Simeone_PRL2023}. The critical density profile corresponds to the saddle point and is determined by solving the Euler-Lagrange equation, namely $\delta \Delta \varOmega^\mathrm{SGA} / \delta \widetilde{\rho} = 0$. Once the critical profile is computed, the critical work of formation is calculated with Eq.~(\ref{SGA}).

\vspace{0.25cm}
\noindent\textit{Results}\,---\,To begin, from Fig.~\ref{fig:rho_c,R_c}(a) which shows representative critical clusters obtained with our rare-event sampling strategy, one can already observe that along with the size, the density of the nucleating cluster seems to decrease when increasing supersaturation. Then, more quantitatively, Figs.~\ref{fig:rho_c,R_c} (b) and \ref{fig:rho_c,R_c}(c) show the critical cluster radii and densities averaged over the whole distribution of clusters obtained in simulations. As predicted by other multidimensional nucleation theories~\cite{Reiss_1976,Schmelzer_2007,Philippe_2011_Phil_mag}, our numerical results confirm that the density of the critical cluster decreases with supersaturation. From the theoretical viewpoint, on the one hand, the critical cluster size is similarly predicted by all models. At the lowest supersaturation, there is a strong agreement with the simulation data. However, all models slightly underestimate the cluster size at intermediate supersaturation levels. On the other hand, CNT fails at capturing the observed behavior, i.e., critical clusters are formed with a density potentially far from that of the equilibrium phase. Meanwhile, although it is able to qualitatively retrieve this effect, the SGA diffuse-interface model~\cite{Cahn-Hilliard-1959,Philippe_JCP_2011,Lutsko-JCP2011,Simeone_PRL2023} overestimates the cluster density and exhibits a maximum that is not seen in simulations. In contrast, our two-variable model successfully reproduces numerical results both qualitatively and quantitatively.

\begin{figure}[!t]
\begin{tikzpicture}
\node (fig1) at (0.4cm,0) {\includegraphics[trim={0cm 1.2cm 13.2cm 0cm},clip,width=3.0cm]{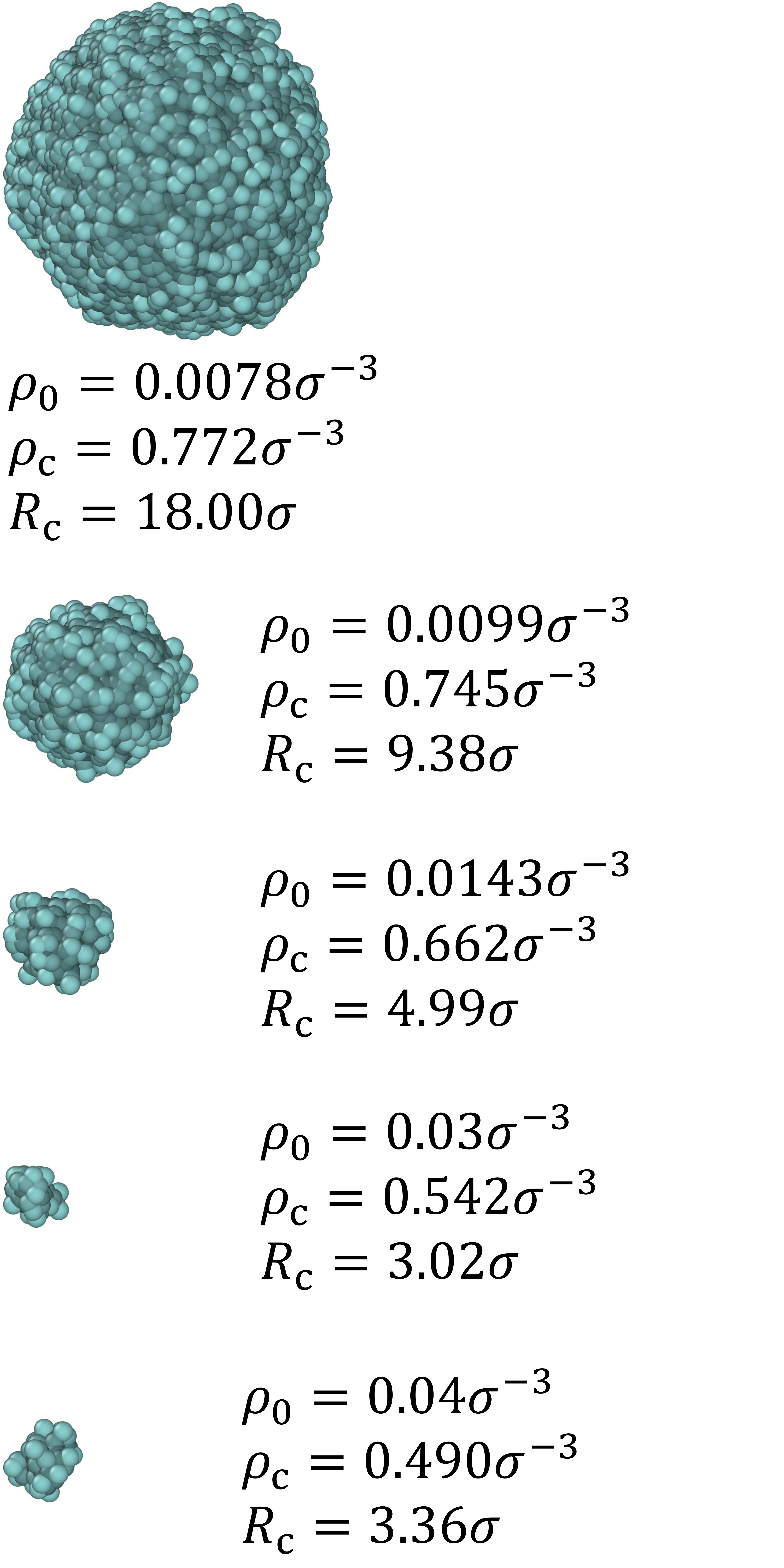}};
\node[] at (-0.9,4) {(a)};
\node (fig2) at (4.5cm,2.1cm) {\includegraphics[trim={6.5cm 11cm 6.7cm 10.2cm},clip,width=5cm]{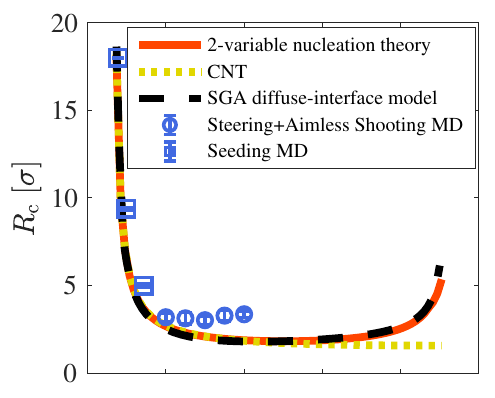}};
\node[] at (3.2,0.6) {(b)};
\node (fig3) at (4.5cm,-2.0cm) {\includegraphics[trim={6.5cm 10.4cm 6.7cm 10.25cm},clip,width=5cm]{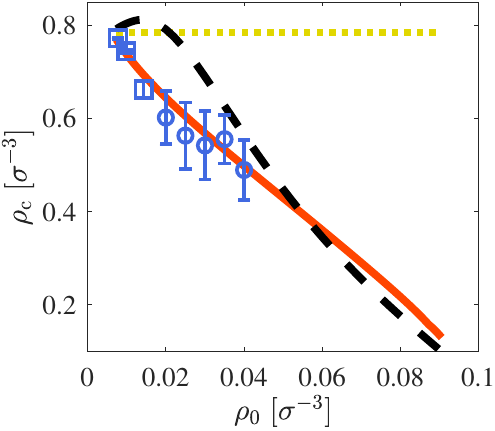}};
\node[] at (3.2,-3.1) {(c)};
\end{tikzpicture}
\caption{(a) Snapshots of critical clusters from our simulations and evolution of the (b) radius and (c) density of critical cluster against initial density at temperature $T=0.8\epsilon/k_\mathrm{B}$. The first three snapshots are from seeding simulations, and the last two are from steering + aimless shooting simulations. Our seeding and aimless shooting MD simulation results are represented by blue squares and circles, respectively, with error bars indicating standard deviations. Predictions from our two-variable nucleation theory, CNT, and the SGA diffuse-interface model are represented by red solid lines, yellow dotted lines, and black dashed lines, respectively.}
\label{fig:rho_c,R_c}
\end{figure}
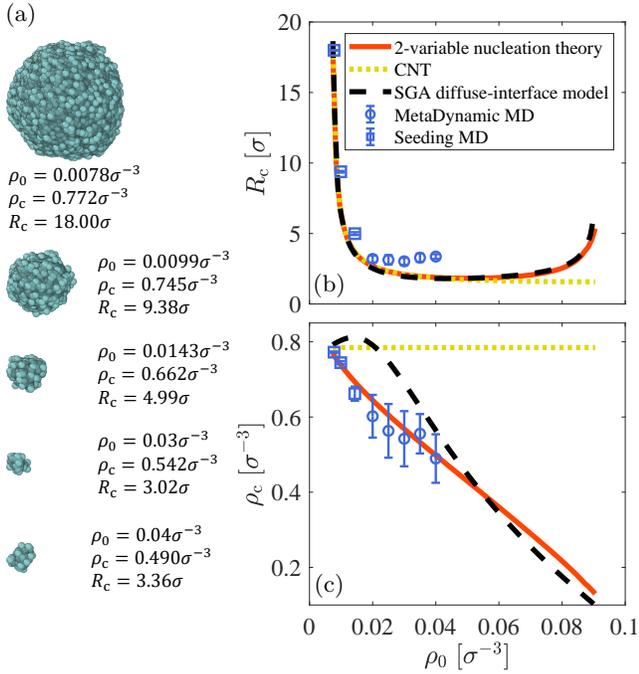

\begin{figure}[!t]
\includegraphics[trim={4.2cm 10.1cm 4.8cm 10.2cm},clip,width=0.45\textwidth]{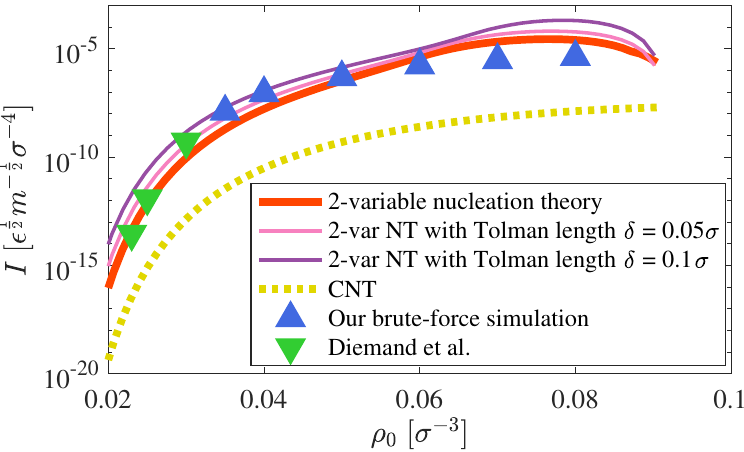}
\caption{Evolution of the steady-state nucleation rate against initial density at $T=0.8\epsilon/k_\mathrm{B}$. Our brute-force simulation results are represented by blue up triangles. Results from Diemand \textit{et al.}~\cite{Diemand_JCP2013} are represented by green down triangles. Predictions from our two-variable nucleation theory and CNT are represented by a red solid line and a yellow dotted line, respectively. Predictions from our two-variable nucleation theory incorporated with Tolman correction of $\delta=0.05\sigma$ and $\delta=0.1\sigma$ are represented by the pink and violet thinner solid lines, respectively. }
\label{fig:I}
\end{figure}

Next, we use our theoretical model and CNT to evaluate the steady-state nucleation rate (see Fig.~\ref{fig:I}). Computing the nucleation rate within the SGA model and comparing it with our two-variable theory would be very valuable since both approaches go beyond the simplistic one-variable CNT. However, coupling the SGA model with the DDFT dynamics to calculate the nucleation rate is not trivial~\cite{Simeone_PRL2023,Lutsko-JCP2024} due to the high dimensionality of diffuse-interface formalisms. This calculation is not performed here. Reducing the problem to two dimensions in our sharp-interface theory, but with the cluster density allowed to vary, results in a straightforward calculation of the nucleation rate. Then, we compare those predictions to numerical results obtained in the literature or using our own brute-force simulations reported in Table I in Appendix and analyzed with the Yasuoka-Matsumoto threshold method~\cite{Yasuoka_Matsumoto_JCP98, Diemand_JCP2013}. Remarkably, the two-variable model performs very well at high supersaturation, where CNT is known to fail by several orders of magnitude. Our model contains no parameters adjusted to nucleation simulations, which demonstrates its predictive capability when properly applied and coupled to quantitative thermodynamics and kinetics data. Furthermore, contrary to CNT, the nucleation rate shows a maximum near the spinodal limit, an effect also observed in diffuse-interface models~\cite{Simeone_PRL2023}. We have so far neglected the Tolman correction~\cite{Tolman_1949}, which accounts for the curvature dependence of surface tension and becomes relevant when the cluster radius is comparable to the interface thickness~\cite{Dellago_PNAS_2016,Azouzi2013,Magaletti2021}. While diffuse-interface theories naturally incorporate this correction~\cite{Prestipino}, it must be explicitly added in sharp-interface models. Here, we implement it by multiplying the surface energy contribution in $\Delta\varOmega$ by $(1-2\delta/R)$, retaining only the first order term, where $\delta$ is the conventional Tolman length~\cite{Tolman_1949,Angelil_JCP2014,Dillmann_1991,Prestipino}. Although the curvature correction has only a minor influence on the critical cluster properties, it noticeably affects the nucleation rate due to its exponential dependency in Eq.~(\ref{eq:I}). Nucleation rates with Tolman correction are shown in Fig.~\ref{fig:I} for $\delta=0.05\sigma$ and $\delta=0.1\sigma$. To preserve good agreement with MD results, our analysis indicates that the Tolman length must remain small, in line with the value $\delta=0.09\sigma$ reported in Ref.~\cite{Angelil_JCP2014} under comparable conditions ($T=0.8\epsilon/k_\mathrm{B}$ and a cutoff of $5\sigma$).

\begin{figure}[!t]
\includegraphics[trim={4.2cm 11.1cm 4.8cm 11.2cm},clip,width=0.45\textwidth]{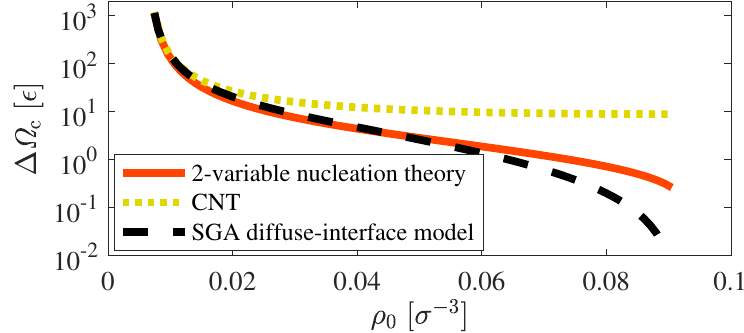}
\caption{Evolution of the critical work of formation against initial density at $T=0.8\epsilon/k_\mathrm{B}$. Predictions from our two-variable nucleation theory, CNT, and the SGA diffuse-interface model are represented by a red solid line, a yellow dotted line, and a black dashed line, respectively.}
\label{fig:DeltaOmega_c}
\end{figure}

Prompted by the confirmations of its quality, our model can be used to predict the work of formation; see Fig.~\ref{fig:DeltaOmega_c}. The Tolman correction, which is almost imperceptible on the nucleation barrier, is not shown. At low supersaturation, the three models agree because critical cluster density matches closely the equilibrium liquid density. In the intermediate supersaturation region, both our model and the SGA model predict much smaller nucleation barriers. It is counterintuitive at first since the driving force for nucleating a dilute cluster is smaller as compared to a cluster with the equilibrium density, but it is balanced by a much lower surface energy cost. Finally, when approaching the spinodal limit, the two-variable nucleation theory recovers results consistent with diffuse-interface models: the density of the critical cluster approaches the initial phase density, its size diverges, and the associated work of formation vanishes. This behavior provides a smooth transition between nucleation and spinodal decomposition, where the energy barrier is zero, as established in diffuse-interface models~\cite{Cahn-Hilliard-1959} and recovered in previous nucleation models with two parameters~\cite{Schmelzer_2000,Schmelzer_2006,Schmelzer_2003,Schmelzer_2011,Baidakov_2000,Philippe_2011_Phil_mag,Philippe_2024,Bonvalet_Phil_mag_2014,Ghosh-2011,Reiss_1976,Schmelzer_2007}. The surface energy contribution also vanishes at the spinodal limit in our two-variable theory since the cluster density approaches that of the initial phase. CNT, however, fails to capture such effects, highlighting the improved physical accuracy of the two-variable model. 

\begin{figure}[b]
\includegraphics[trim={4.3cm 9.3cm 4.3cm 9.2cm},clip,width=0.45\textwidth] {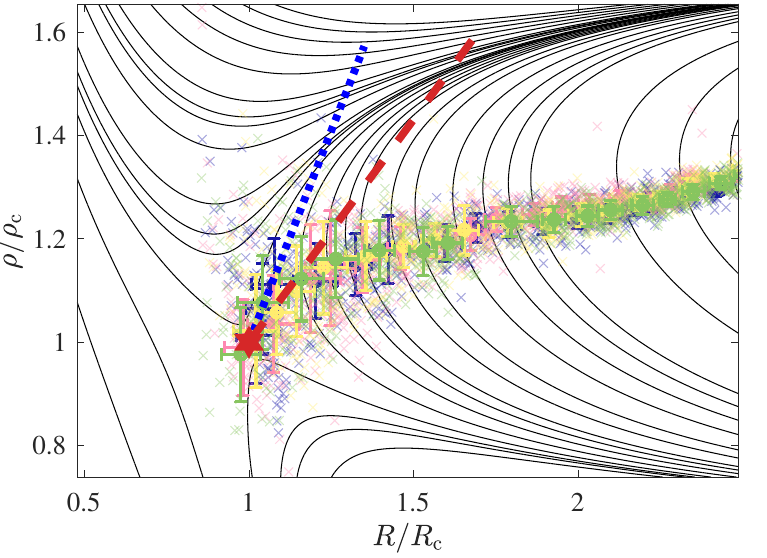}
\caption{Contour plot of the work of formation overlaid with growth trajectories of critical clusters from simulations at $T=0.8\epsilon/k_\mathrm{B}$ and $\rho_0=0.03\sigma^{-3}$. The work of formation is computed using our two-variable nucleation theory (without Tolman correction), with the saddle point marked by a red hexagram. The red dashed line and blue dotted line indicate the predicted nucleation flux direction and steepest decent direction at the saddle point, respectively. Growth trajectories from four independent simulations are represented by colored crosses, with corresponding mean values shown as dots. Error bars denote standard deviations. $(R,\rho)$ coordinates from theory and simulation are normalized by their corresponding critical values. }
\label{fig:WorkOfFormation_Trajectories}
\end{figure}

Finally, we investigate the nucleation flux direction. In CNT, both kinetics and thermodynamics favor cluster growth at equilibrium density. However, in our two-variable theory the nucleation flux is found to deviate from the energy steepest descent direction when supersaturation increases (see Fig.~\ref{fig:WorkOfFormation_Trajectories}). Four growth trajectories obtained from brute-force simulations are also superimposed to the corresponding free energy surface at $\rho_0=0.03\sigma^{-3}$. These simulations are initialized with the critical cluster shown in Fig.~\ref{fig:rho_c,R_c}(a), with therefore an equal probability ($\sim 0.5$) of growing or shrinking but we keep only growth trajectories for the sake of clarity. From Fig.~\ref{fig:WorkOfFormation_Trajectories}, we observe that contrary to the CNT predictions, the nucleation pathway is not an horizontal line but involves concomitant growth and densification of the cluster. More importantly, the numerical results align more closely with our model's prediction than with the energy steepest descent direction, providing further confirmation of the quality of the two-variable nucleation theory.  

\vspace{0.25cm}
\noindent\textit{Conclusions}\,---\,We revisited the longstanding yet still fundamental problem of liquid condensation and uncovered a nonclassical nucleation pathway. Using innovative rare-event simulation strategies---an original implementation of the seeding approach together with steering MD combined with aimless shooting---we were able to probe nucleation events with unprecedented precision and collect structures situated at the saddle point of the free energy barrier across a wide range of supersaturation. Our simulation results show that the density of the critical clusters differs markedly from that of the equilibrium liquid and decreases as the vapor density increases. Building on this key insight, we expanded the capillary model by explicitly incorporating the cluster density as a supplementary driving parameter. Without any fitted parameters, this self-consistent two-variable nucleation theory successfully captures the concomitant growth and densification of nearly critical clusters, as observed in our simulations, and also achieves quantitative agreement for nucleation rates and critical cluster densities. Taken together, these results demonstrate that cluster density is an essential collective variable governing nucleation, and that omitting it is a principal reason for the longstanding failures of CNT. By establishing a multidimensional picture of nucleation, our work provides a broadly applicable framework for refining nucleation theories.

\vspace{0.25cm}
\begin{acknowledgments}
\noindent\textit{Acknowledgments}\,---\,This work was supported by the ANR TITANS project. Computational resources have been provided by the DSI at Université de Lille, Jean-Zay and TGCC.

\end{acknowledgments}

\bibliography{reference}

\appendix

\newpage
\clearpage

\begin{center}
\Huge{Appendix}
\end{center}

\section{Simulation protocol}

Simulations were performed using the open-source package \texttt{LAMMPS} \cite{LAMMPS}. For the Lennard-Jones interactions, we used a large cutoff of $6.78\sigma$, ensuring excellent agreement between the phase diagram obtained from perturbation theory and that from simulations \cite{baidakov_self-diffusion_2011,baidakov_metastable_2007}. The employed timestep is $0.005\tau$, with $\tau=\sigma m^{\frac{1}{2}} \epsilon^{-\frac{1}{2}}$. Specifically, we work at $T=0.8\epsilon/k_\mathrm{B}$. For temperature and pressure controls, we used the Nose-Hoover thermostat/barostat implemented in \texttt{LAMMPS} with damping parameters taken respectively as $0.5\tau$ and $5\tau$.

At low supersaturation, the \emph{NVT} seeding technique is employed to stabilize liquid clusters in equilibrium with the vapor \cite{Vega-Seeding-NVT}. Once the cluster is stabilized, the simulation is converted to the \emph{NPT} ensemble to emulate an infinite system. In this stage, the pressure is prescribed as the time-averaged virial pressure from the preceding \emph{NVT} simulation \cite{PhysRevE.101.022611}. The system volume becomes a free variable, accommodating itself to maintain the pressure. As the considered cluster is critical (but now unstable), it has an equal probability ($\sim 0.5$) of growing, from which growth rates can be estimated, or shrinking. To identify critical clusters and their growth rates, multiple \emph{NVT}-to-\emph{NPT} runs are performed from different initial configurations, varying particle positions and velocities while keeping the desired quantities. In the \emph{NVT} ensemble, we were able to stabilize 20 liquid clusters per box size ($20\sigma$, $50\sigma$ and $90\sigma$). When checking their criticality in infinite systems, 20 \emph{NPT} runs were performed.

In intermediate supersaturation regimes, we used a set of biasing techniques. Here are some details on the implementation. Firstly, for the steered MD, simulations were run for $10^4 \tau$ with $10^4$ atoms. For the collective variable, we used $N_\mathrm{liq}$ defined as the number of atoms having coordination number larger than 6. The constraint potential constant was chosen as $100/s^2$ where $s$ is the standard deviation of $N_\mathrm{liq}$ measured in the vapor phase. Steered MD were performed three times for each of the considered vapor density to ensure repeatability. Secondly, from the steered MD accumulated structures, we measured the commitment probability to obtain brute-force trajectory connecting vapor and liquid phases using 10 different initial velocity conditions. For each of the three steered MD simulations, we obtained one connecting trajectory. Thirdly, the obtained trajectories were used as initialization for  three-point aimless shooting algorithm\,\cite{Peters2006Aug} where we sampled configurations separated by a time interval equal to $10\tau$. From each of the connecting trajectory, we launched 10 different aimless shooting sampling. Finally, we measured the commitment probability of every structures obtained with aimless shooting using 50 initial velocities. Structures were considered as critical when there commitment probability is located in the range $[0.3,0.7]$. More details can be found in Ref. \cite{Lam2024}.

At high supersaturation, brute-force simulations were performed in the \emph{NVT} ensemble, in boxes of size $L=100\sigma$. 5 independent simulations were launched for each initial density. This change in system sizes allows us to obtain many nucleation events at the same time thus matching with the Yasuoka-Matsumoto method.

\section{Simulation analysis}
Cluster radius $R$, density $\rho$, and vapor density $\rho_0$ from simulations were determined by fitting the integral of a modified sigmoid function \cite{Lutsko_2008},
\begin{equation*}
    \widetilde{\rho}(r)=\rho_0+(\rho-\rho_0)\frac{1+br}{1+(br)^2}\frac{1-\tanh(A(r-R))}{1-\tanh(-AR)},
\end{equation*}
where $A$ represents the liquid/vapor interface width and $b=2A/(\exp{(2AR)}+1)$, to the integrated radial density profile calculated from the center of the stable cluster. Post-processing was implemented in \texttt{Python}, utilizing \texttt{Ovito}'s MD analysis library \cite{ovito}.

\section{Quantitative outputs from the simulation}
Table \ref{tab:SimulationResults} shows the critical densities, radii, and steady-state nucleation rates measured from our MD simulations. 
\begin{table}[h!]
\caption{Critical cluster density, radius, and nucleation rate obtained from our MD simulations at $T=0.8\epsilon/k_\mathrm{B}$.}
\begin{ruledtabular}
\begin{tabular}{lccc}
$\rho_0\,[\sigma^{-3}]$ & $\rho_\mathrm{c}\,[\sigma^{-3}]$ & $R_\mathrm{c}\,[\sigma]$ & $I\,[\epsilon^{\frac{1}{2}} m^{-\frac{1}{2}} \sigma^{-4}]$ \\
\colrule
$0.0078$ & $0.772\pm0.002$ & $18.00\pm0.02$ & \\
$0.0099$ & $0.745\pm0.004$ & $9.38\pm0.06$ & \\
$0.0143$ & $0.662\pm0.019$ & $4.99\pm0.10$ & \\
$0.02$ & $0.602\pm0.057$ & $3.19\pm0.25$ & \\
$0.025$ & $0.563\pm0.071$ & $3.13\pm0.33$ & \\
$0.03$ & $0.542\pm0.073$ & $3.02\pm0.25$ & \\
$0.035$ & $0.556\pm0.052$ & $3.28\pm0.29$ & $1.3\pm0.3\times10^{-8}$ \\
$0.04$ & $0.490\pm0.065$ & $3.36\pm0.17$ & $9.1\pm0.8\times10^{-8}$ \\
$0.05$ &   &   & $4.9\pm1.2\times10^{-7}$ \\
$0.06$ &   &   & $1.7\pm0.3\times10^{-6}$ \\
$0.07$ &   &   & $3.0\pm0.4\times10^{-6}$ \\
$0.08$ &   &   & $4.3\pm0.7\times10^{-6}$ \\
\end{tabular}
\end{ruledtabular}
\label{tab:SimulationResults}
\end{table}

\section{Classical nucleation theory (CNT) predictions}
\label{appendix:CNT}
The predictions from CNT are presented in the figures. Here we outline the corresponding formulae.
The critical cluster density is assumed to be equal to the equilibrium liquid density, 
\begin{equation*}
    \rho_\mathrm{c}^\mathrm{CNT} = \rho_2^\mathrm{eq}.
\end{equation*}
As a result, the driving force of nucleation is given by $g_n^\mathrm{CNT} = g_n(\rho_2^\mathrm{eq}) $. 
The surface tension is assumed to be the one at equilibrium, $\gamma^\mathrm{eq}$. Consequently, for a given initial density, the work of formation depends only on the cluster radius $R$: 
\begin{equation*}
    \Delta\varOmega^\mathrm{CNT}(R)=-\frac{4}{3}\mathrm{\pi}{R}^3g_n^\mathrm{CNT}+ 4\mathrm{\pi}{R}^2\gamma^\mathrm{eq}.
\end{equation*}
The critical cluster radius and the critical work of formation are given by
\begin{equation*}
    R_\mathrm{c}^\mathrm{CNT} = \frac{2\gamma^\mathrm{eq}}{g_n^\mathrm{CNT}}
    \quad \text{and} \quad
    \Delta\varOmega_\mathrm{c}^\mathrm{CNT} = \frac{16\mathrm{\pi}(\gamma^\mathrm{eq})^3}{3(g_n^\mathrm{CNT})^2}.
\end{equation*}
The Hessian matrix $\mathbf{H}$ reduces to a scalar: 
\begin{equation*}
    \mathrm{H^{CNT}} = -8\mathrm{\pi}\gamma^\mathrm{eq}.
\end{equation*}

Kinetics is calculated from DDFT at the critical point. $\mathbf{D}$ reduces into a scalar as well:  
\begin{equation*}
    \mathrm{D^{CNT}} = D_\mathrm{diff}\frac{\rho_0}{4\mathrm{\pi} (\rho_\mathrm{c}^\mathrm{CNT}-\rho_0)^2 (R^\mathrm{CNT}_\mathrm{c})^3}.
\end{equation*}

The distribution parameter is computed from mass conservation,
\begin{equation*}
    f_0^\mathrm{CNT} = \frac{\rho_0} {\frac{4}{3}\mathrm{\pi} \rho_\mathrm{c}^\mathrm{CNT} \displaystyle\int_0^{R_\mathrm{c}^\mathrm{CNT}} R^3 \exp{(-\beta\Delta\varOmega^\mathrm{CNT}) \,\mathrm{d}R}} .
\end{equation*}

Finally, the nucleation rate predicted by CNT is given by \cite{alekseechkin_multivariable_2006, langer_statistical_1969}
\begin{equation*}
    I^\mathrm{CNT} =  f_0^\mathrm{CNT} \cdot \sqrt{\dfrac{-\beta \mathrm{H^{CNT}}}{2\mathrm{\pi}}} \cdot \mathrm{D^{CNT}} \cdot \exp(-\beta \Delta\varOmega_\mathrm{c}^\mathrm{CNT}).
\end{equation*}

\section{Incorporation of dynamical density functional theory (DDFT) into nucleation theory}
\label{appendix:DDFT}
In DDFT, the driving force for particle diffusion is assumed to be proportional to the gradient of the chemical potential in the real space $(\mathbf{r})$ \cite{Evans01041979},
\begin{equation*}
    \boldsymbol{j} (\mathbf{r}) = -\beta D_\mathrm{diff}\widetilde{\rho}(\mathbf{r}) \boldsymbol{\nabla} \mu(\mathbf{r}),
\end{equation*}
where $D_\mathrm{diff}$ is the diffusion constant, $\widetilde{\rho}(\mathbf{r})$ is the real-space density profile and $\mu(\mathbf{r}) = \frac{\delta F [\widetilde{\rho}(\mathbf{r})]} {\delta\widetilde{\rho}(\mathbf{r})}$ is the chemical potential of the system. It is important to note that $F$ represents the total Helmholtz free energy of the system, whereas $\mathcal{F}$ in the main text denotes the Helmholtz free energy per unit volume of a homogeneous phase. 

Combined with the continuity equation,
\begin{equation*}
    \frac{\partial\widetilde{\rho}(\mathbf{r})}{\partial t} = - \boldsymbol{\nabla\cdot} \boldsymbol{j} (\mathbf{r}), 
\end{equation*}
the evolution of the time-dependent density profile is governed by  
\begin{equation}
    \frac{\partial \widetilde{\rho}(\mathbf{r})}{\partial t} = \beta D_\mathrm{diff}\boldsymbol{\nabla\cdot}\Bigg(\widetilde{\rho}(\mathbf{r})\boldsymbol{\nabla} \frac{\delta F[\widetilde{\rho}(\mathbf{r})]}{\delta \widetilde{\rho}(\mathbf{r})}\Bigg) .
\label{eq:DDFT}
\end{equation}

Following closely the method proposed by Lutsko \cite{lutsko_communication_2011}, we parameterize the density profile in order to derive a dynamical equation in the phase space $\{\boldsymbol{x}\}$ for the nucleation theory. We define $n(r,t)$ as the integrated density within a spherical shell of radius $r$,
\begin{equation}
    n(r,t) = \int_{r'<r} \widetilde{\rho}(\mathbf{r'},t)\, \mathrm{d}\mathbf{r'} .
\label{eq:n_DDFT}
\end{equation}
Assuming spherical symmetry, we integrate Eq. (\ref{eq:DDFT}) as Eq. (\ref{eq:n_DDFT}) and apply Gauss' theorem, obtaining 
\begin{equation}
    \dfrac{1}{4\mathrm{\pi} r^2\widetilde{\rho}(r)} \cdot \frac{\partial n(r)}{\partial t} = \beta D_\mathrm{diff}\frac{\partial \mu(r)}{\partial r} ,
\label{eq:DDFT1}
\end{equation}
where $r$ is the radial coordinate. 

The density profile is then parameterized as $\widetilde{\rho}(\mathbf{r},t)\longmapsto\widetilde{\rho}(\mathbf{r},\boldsymbol{x}(t))$, if the variables $\boldsymbol{x}$ are capable of fully describing the density profile under the assumed model. For instance, in the two-variable framework, the density profile is approximated as a step function:  
\begin{equation*}
    \widetilde{\rho}(r)=\left\{
    \begin{aligned}
        &\rho, \,\quad r<R    \\[7pt]
        &\rho_0, \quad r>R
    \end{aligned}
    \right.
    ,
\end{equation*}
where $R$ and $\rho$ are sufficient to describe it. Consequently, the particle number $n(r,t)$ is rewritten as $n(r,\boldsymbol{x}(t))$, leading to   
\begin{equation*}
    \frac{\partial n}{\partial t} = (\frac{\partial n}{\partial \boldsymbol{x}})^\mathrm{T} \frac{\partial \boldsymbol{x}}{\partial t} .
\end{equation*}
Substituting this expression into Eq. (\ref{eq:DDFT1}), pre-multiplying by $\frac{\partial n}{\partial\boldsymbol{x}}$, and integrating over $r$ from $0$ to infinity, we obtain
\begin{equation}
    \mathbf{g}\frac{\partial\boldsymbol{x}}{\partial t} = \beta D_\mathrm{diff} \int_0^\infty \frac{\partial\mu(r)}{\partial r} \frac{\partial n}{\partial \boldsymbol{x}} \,\mathrm{d}r ,
\label{eq:DDFT_g}
\end{equation}
where the elements of the matrix $\mathbf{g}$ are given by 
\begin{equation}
    \mathrm{g}_{ij} = \int_0^\infty \dfrac{1}{4\mathrm{\pi} r^2\widetilde{\rho}(r)} \cdot \frac{\partial n}{\partial x_i} \cdot \frac{\partial n}{\partial x_j} \,\mathrm{d}r .
\label{eq:gij_DDFT}
\end{equation}
This is the $\mathbf{g}$ matrix introduced by Lutsko \cite{lutsko_communication_2011} in its so-called mesoscopic theory of nucleation. $\mathbf{g}$ can be directly connected to Alekseechkin's theory \cite{alekseechkin_multivariable_2006}. 
The right-hand side of Eq. (\ref{eq:DDFT_g}) can be evaluated using integration by parts, yielding
\begin{equation*}
\begin{aligned}
    \int_0^\infty \frac{\partial\mu(r)}{\partial r} \frac{\partial n}{\partial\boldsymbol{x}} \,\mathrm{d}r 
    = & 
    \frac{\partial n}{\partial \boldsymbol{x}}\mu(r)\bigg|_{r=\infty} - \frac{\partial n}{\partial \boldsymbol{x}}\mu(r)\bigg|_{r=0} \\
    & - \int_0^\infty \frac{\delta \mathcal{F}[\widetilde{\rho}(r)]}{\delta \widetilde{\rho}(r)} \frac{\partial \widetilde{\rho}(r)}{\partial \boldsymbol{x}} \,\mathrm{d}r .
\end{aligned} 
\end{equation*}
At $r=\infty$, the system reaches the homogeneous initial state with chemical potential $\mu_\mathrm{i}$, and its total particle number is $n_\mathrm{total}$. Therefore, the first term of the above equation simplifies to $\frac{\partial}{\partial \boldsymbol{x}} (n_\mathrm{total}\mu_\mathrm{i})$. The second term vanishes because at $r=0$, $\frac{\partial n}{\partial \boldsymbol{x}} = \frac{4}{3}\mathrm{\pi} r^3 \frac{\partial \widetilde{\rho}(r)}{\partial \boldsymbol{x}}$, and $\frac{\partial \widetilde{\rho}(r)}{\partial \boldsymbol{x}}$ remains finite. The third term can be handled using the chain rule of derivative, yielding $\frac{\partial F}{\partial \boldsymbol{x}}$. 
Thus, we obtain 
\begin{equation*}
    \int_0^\infty \frac{\partial \mu(r)}{\partial r} \frac{\partial n}{\partial \boldsymbol{x}} \,\mathrm{d}r = \frac{\partial (n_\mathrm{total}\mu_\mathrm{i})}{\partial \boldsymbol{x}} - \frac{\partial F}{\partial \boldsymbol{x}} = -\frac{\partial \Delta\varOmega}{\partial \boldsymbol{x}} ,
\end{equation*}
where $\Delta\varOmega = F - n_\mathrm{total}\mu_\mathrm{i}$ is indeed the work of formation. Substituting this result into Eq. (\ref{eq:DDFT_g}), we get
\begin{equation*}
    \mathbf{g}\frac{\partial \boldsymbol{x}}{\partial t} = -\beta D_\mathrm{diff}\frac{\partial \Delta\varOmega}{\partial \boldsymbol{x}} .
\end{equation*}
Comparing this with the Fokker-Planck equation at equilibrium, 
\begin{equation*}
    \frac{\partial \boldsymbol{x}}{\partial t} = -\beta \mathbf{D}\frac{\partial \Delta\varOmega}{\partial \boldsymbol{x}}, 
\end{equation*}
we can directly identify the diffusivity matrix as $\mathbf{D}=D_\mathrm{diff}\,\mathbf{g^\mathrm{-1}}$. 
For the two-variable model, the diffusivity matrix is given by \cite{lutsko_two-parameter_2015}:
\begin{equation}
    \mathbf{D} = D_\mathrm{diff}
    \begin{pmatrix}
        4\mathrm{\pi} R^3 \dfrac{(\rho-\rho_0)^2}{\rho_0}  & 4\mathrm{\pi} R^4  \dfrac{\rho-\rho_0}{3\rho_0}\\
        4\mathrm{\pi} R^4  \dfrac{\rho-\rho_0}{3\rho_0} & \frac{4}{9}\mathrm{\pi} R^5 \big(\frac{1}{5\rho}+\frac{1}{\rho_0}\big)
    \end{pmatrix}
    ^{-1}.
\label{eq:D}
\end{equation}


\end{document}